# BAROTROPIC MAGNETOHYDRODYNAMICS AS A FOUR FUNCTION FIELD THEORY WITH NON-TRIVIAL TOPOLOGY AND AHARONOV-BOHM EFFECTS

Asher Yahalom
Ariel University Center of Samaria, Ariel 40700, Israel
E-mail: asya@ariel.ac.il

## ABSTRACT

Variational principles for magnetohydrodynamics were introduced by previous authors both in Lagrangian and Eulerian form. In previous works [1] Yahalom & Lynden-Bell and later Yahalom [2] introduced a simpler Eulerian variational principle from which all the relevant equations of Magnetohydrodynamics can be derived. The variational principles were given in terms of four independent functions for non-stationary flows and three independent functions for stationary flows. This is less than the seven variables which appear in the standard equations of magnetohydrodynamics which are the magnetic field $\vec{B}$, the velocity field $\vec{v}$ and the density $\rho$. In the case that the magnetohydrodynamic flow has a non trivial topology such as when the magnetic lines are knotted or magnetic and stream lines are knotted, some of the functions appearing in the Lagrangian are non-single valued. Those functions play the same rule as the phase in the Aharonov-Bohm celebrated effect [3].

## INTRODUCTION

Variational principles for magnetohydrodynamics were introduced by previous authors both in Lagrangian and Eulerian form. Following the work of Eckart [4] for non-magnetic flows, Newcomb [5] has introduced in his paper a Lagrangian variational formalism for magnetohydrodynamics. A similar formalism was discussed in Sturrock's book [6]. Eulerian variational principles for non-magnetic fluid dynamics were first introduced by Davydov [7]. Following the work of Davydov, Zakharov and Kuznetsov [8] suggested an Eulerian variational principle for magnetohydrodynamics. However, the variational principle suggested by Zakharov and Kuznetsov contained **two** more functions than the standard formulation of magnetohydrodynamics with a total sum of **nine** variational variables. Another Eulerian variational principle for magnetohydrodynamics was introduced independently by Calkin [9] in a work that preceded Zakharov and Kuznetsov paper by seven years. However, Calkin's variational principle also depends on as much as eleven variational variables. The situation was somewhat improved when Vladimirov and Moffatt [10] in a series of papers have discussed an Eulerian variational principle for incompressible magnetohydrodynamics. Their variational principle contained only three more functions in addition to the seven variables which appear in the standard equations of magnetohydrodynamics which are the magnetic field $\vec{B}$ the velocity field $\vec{v}$ and the density $\rho$. Kats [11] has generalized Moffatt's work for compressible non barotropic flows but without reducing the number of functions and the computational load. Sakurai [12] has introduced a two function Eulerian variational principle for force-free magnetohydrodynamics and used it as a basis of a numerical scheme. Yahalom & Lynden-Bell [1, 13] have combined the Lagrangian of Sturrock [6] with the Lagrangian of Sakurai [12] to obtain an Eulerian variational principle depending on only six functions. The vanishing of the variational derivatives of this Lagrangian entail all the equations needed to describe barotropic magnetohydrodynamics without any additional constraints. The equations obtained resemble the Hamiltonian equations of Frenkel, Levich & Stilman [14] (see also [15]), the same Hamiltonian equations were obtained at around the same time independently by Morrison [16] who was concerned about obtaining proper Poisson brackets for magnetohydrodynamics. Furthermore, it was shown by Yahalom & Lynden-Bell [1] that for stationary flows three functions will suffice in order to describe a Lagrangian principle for barotropic magnetohydrodynamics. Later Yahalom [2] has shown that the number of functions needed to describe magnetohydrodynamics can be reduced further and that indeed four functions suffice in the case of non-stationary flows.

The non-singlevaluedness of the functions appearing in the reduced representation of barotropic magnetohydrodynamics was discussed in particular with connection to the topological invariants of magnetic and cross helicities. It was shown that flows with non trivial topologies which have non zero magnetic or cross helicities can be adequately described by the functions of the reduced representation provided that some of them are non-single valued [1, 13]. The cross helicity per unit flux was shown to be equal to the discontinuity of the function $\nu$, this discontinuity was shown to be a conserved quantity along the flow. The magnetic helicity per unit flux was shown to be equal to the discontinuity of another function $\zeta$. It should be mentioned that the existence of non single valued functions in the description of toroidal magnetohydrodynamics was first suggested by Kruskal & Kulsrud [17].

Aharonov and Bohm [3] have shown that a confined magnetic field will effect the trajectory of an electron even if the electron is restricted to move only in a domain where the magnetic field is null, this was verified experimentally and was thought to be a victory of quantum mechanics over classical mechanics were such effects are not supposed to exist. In this paper we will show that Aharonov and Bohm effect is a topological effect and that an analogue topological effects exist in classical continuum mechanics in particular in magnetohydrodynamics. Thus the phase of the Aharonov and Bohm is non single valued for the same reason that the functions $\nu$ and $\zeta$ are not single valued.

The plan of this paper is as follows: First I introduce the salient features of the Aharonov and Bohm [3] effect, then I introduce the standard notations and equations of barotropic magnetohydrodynamics. Next I introduce the functions needed to describe the Lagrangian and the variational formalism follows. Finally I discuss magnetohydrodynamics with non-trivial topology and its relations to the Aharonov-Bohm effect.

**AHARONOV-BOHM EFFECT**

Consider an electron moving from A to B (figure 1) in the middle we have a magnetic field $\vec{B}$ going into the plane through which the electron is forbidden to pass, hence for the electron the magnetic field is zero. However, the vector potential $\vec{A}$ is not zero, in fact:

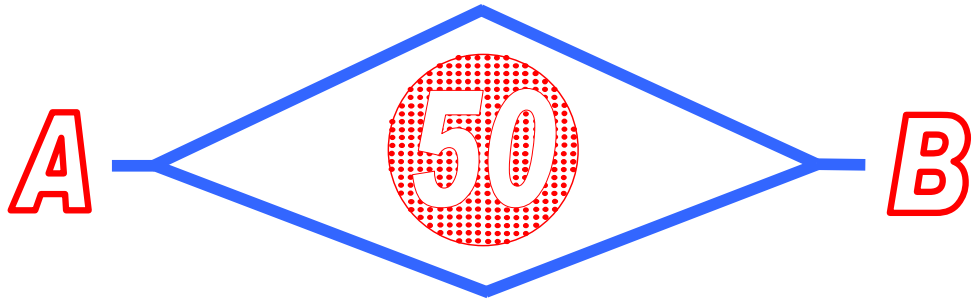

Figure **1**: An electron is moving from point A to point B. In the middle of the figure we have a confined magnetic field of 50 Tesla.
(Alternatively A and B are the initials of Aharonov and Bohm and 50 is the time passed from the discovery of the effect till the year 2009)

$$\vec{B} = \vec{\nabla} \times \vec{A} = 0 \Rightarrow \vec{A} = \vec{\nabla}\overline{S} \qquad (1)$$

$\vec{\nabla}$ has its standard meaning in vector calculus, $\overline{S}$ is a non single valued function and its discontinuity $[\overline{S}]$ can be calculated immediately using Stokes theorem:

$$\Phi = \int \vec{B} \cdot d\vec{S} = \int \vec{\nabla} \times \vec{A} \cdot d\vec{S} = \oint \vec{A} \cdot d\vec{l} = \oint \vec{\nabla}\overline{S} \cdot d\vec{l} = [\overline{S}] \qquad (2)$$

Here $\Phi$ is the magnetic flux, the first integral is an area integral and the third is a line integral in which the trajectory goes around the confined magnetic field. Aharonov and Bohm [3] have shown that $\overline{S}$ is proportional to the phase of the electron wave function. Thus its discontinuity will cause interference at point B. If the magnetic field is uniform in a cylinder and zero outside the cylinder, the vector potential can be calculated to be:

$$\vec{A} = A_\theta \hat{\theta} = \frac{\Phi}{2\pi r}\hat{\theta} = \vec{\nabla}\overline{S} \Rightarrow \overline{S} = \frac{\Phi}{2\pi}\theta + \overline{S}_0. \qquad (3)$$

Where $\theta$ is the azimuthal angle.

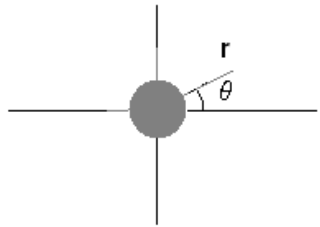

Figure **2**: The azimuth-radius coordinate system, used to calculate the vector potential. The magnetic field vanishes except at the gray area.

The main features of the Aharanov - Bohm effect are:

1. A domain that is not simply connected due to the presnce of a magnetic field, but can be made simply connected by introducing a cut. Mathematically speaking the domain has a non-trivial fundamental Homotopy group. Two classes of loops exist in the plain, loops that can be contracted to a point without intersecting the magnetic region and loops that can not.
2. The electron (or its wave function) do not feel directly the magnetic field – non locality.
3. The potential vector field is a gradient of a non-single valued function.
4. Gauge freedom is not gone but only limited to single-valued gauges.

To conclude we mention that according to Bohm's causal interpretation of quantum mechanics there is a quantum - classical correspondence. According to Bohm [18,19] the phase of a wave function $S$ should be interpreted as a potential of the velocity field $\vec{v}$:

$$\vec{v} = \frac{1}{m}\vec{\nabla}S \qquad (4)$$

$m$ is the mass of the particle. However, this correspondence can go the other way around!! If the velocity field has a potential part it can be interpreted as a phase of a wave function.

**STANDARD FORMULATION OF MAGNETOHYDRODYNAMICS**

The standard set of equations solved for barotropic magnetohydrodynamics are given below:

$$\frac{\partial \vec{B}}{\partial t} = \vec{\nabla} \times (\vec{v} \times \vec{B}), \qquad (5)$$

$$\vec{\nabla} \cdot \vec{B} = 0, \qquad (6)$$

$$\frac{\partial \rho}{\partial t} + \vec{\nabla} \cdot (\rho \vec{v}) = 0, \qquad (7)$$

$$\rho \frac{d\vec{v}}{dt} = \rho(\frac{\partial \vec{v}}{\partial t} + (\vec{v} \cdot \vec{\nabla})\vec{v}) = -\vec{\nabla}p(\rho) + \frac{(\vec{\nabla} \times \vec{B}) \times \vec{B}}{4\pi}. \qquad (8)$$

The following notations are utilized: $\frac{\partial}{\partial t}$ is the temporal derivative, $\frac{d}{dt}$ is the temporal material derivative and $\rho$ is the fluid density. Finally $p(\rho)$ is the pressure which we assume depends on the density alone (barotropic case). The justification for those equations and the conditions under which they apply can be found in standard books on magnetohydrodynamics (see for example [6]). Equation (5) describes the fact that the magnetic field lines are moving with the fluid elements ("frozen" magnetic field lines), equation (6) describes the fact that the magnetic field is solenoidal, equation (7) describes the conservation of mass and equation (8) is the vector Euler equation for a fluid in which both pressure and Lorentz magnetic forces apply. The term:

$$\vec{J} = \frac{\vec{\nabla} \times \vec{B}}{4\pi}, \qquad (9)$$

is the electric current density which is not connected to any mass flow. The number of independent variables for which one needs to solve is seven ($\vec{v}, \vec{B}, \rho$) and the number of equations (5,7,8) is also seven. Notice that equation (6) is a condition on the initial $\vec{B}$ field and is satisfied automatically for any other time due to equation (5). Also notice that $p(\rho)$ is not a variable rather it is a given function of $\rho$.

## POTENTIAL REPRESENTATION OF VECTOR QUANTITIES OF MAGNETOHYDRODYNAMICS

It was shown in [1,2] that $\vec{B}$ and $\vec{v}$ can be represented in terms of three scalar functions $\chi, \eta, \nu$. Following Sakurai [12] the magnetic field takes the form:

$$\vec{B} = \vec{\nabla}\chi \times \vec{\nabla}\eta. \qquad (10)$$

Hence $\vec{B}$ satisfies automatically equations (5,6) for co-moving $\chi$ and $\eta$ surfaces and is orthogonal to both $\vec{\nabla}\chi$ and $\vec{\nabla}\eta$. The above expression can also describe a magnetic field with non-zero magnetic helicity as was demonstrated in [1]. Moreover, the velocity $\vec{v}$ can be represented in the following form [2]:

$$\vec{v} = \vec{\nabla}\nu + \frac{1}{\vec{B}^2}[\frac{\partial \eta}{\partial t}\vec{\nabla}\chi - \frac{\partial \chi}{\partial t}\vec{\nabla}\eta + \vec{\nabla}\nu \times \vec{B}] \times \vec{B}$$
$$= \frac{1}{\vec{B}^2}[(\frac{\partial \eta}{\partial t}\vec{\nabla}\chi - \frac{\partial \chi}{\partial t}\vec{\nabla}\eta) \times \vec{B} + \vec{B}(\vec{\nabla}\nu \cdot \vec{B})] \equiv \vec{v}_\perp + \vec{v}_\parallel \qquad (11)$$

Hence the velocity $\vec{v}$ is partitioned naturally into two components one which is parallel to the magnetic field and another one which is perpendicular to it. This choice of $\vec{v}$ assures us the that $\chi, \eta$ are indeed co-moving.

## THE ACTION OF BAROTROPIC MAGNETOHYDRODYNAMICS

The Lagrangian density of barotropic magnetohydrodynamics was shown to depend on four functions $\chi, \eta, \nu, \rho$ and to take the form [2]:

$$L[\chi, \eta, \nu, \rho] = \rho[\frac{1}{2}\vec{v}^2 - \frac{d\nu}{dt} - \varepsilon(\rho)] - \frac{\vec{B}^2}{8\pi} \qquad (12)$$

where $\vec{v}$ is given by equation (11), $\vec{B}$ by equation (10) and $\varepsilon(\rho)$ is the internal energy density. Or more explicitly as:

$$L[\chi, \eta, \nu, \rho] = \frac{1}{2}\frac{\rho}{(\vec{\nabla}\chi \times \vec{\nabla}\eta)^2}[\vec{\nabla}\eta\frac{\partial \chi}{\partial t} - \vec{\nabla}\chi\frac{\partial \eta}{\partial t} + (\vec{\nabla}\chi \times \vec{\nabla}\eta) \times \vec{\nabla}\nu]^2$$
$$- \rho[\frac{\partial \nu}{\partial t} + \frac{1}{2}(\vec{\nabla}\nu)^2 + \varepsilon(\rho)] - \frac{(\vec{\nabla}\chi \times \vec{\nabla}\eta)^2}{8\pi}. \qquad (13)$$

This Lagrangian density admits an infinite symmetry group of transformations of the form:

$$\hat{\eta} = \hat{\eta}(\chi,\eta), \qquad \hat{\chi} = \hat{\chi}(\chi,\eta), \tag{14}$$

provided that the absolute value of the Jacobian of these transformation is unity:

$$\left|\frac{\partial(\hat{\eta},\hat{\chi})}{\partial(\eta,\chi)}\right| = 1. \tag{15}$$

In particular the Lagrangian density admits an exchange symmetry:

$$\hat{\eta} = \chi, \qquad \hat{\chi} = \eta. \tag{16}$$

Taking the variational derivatives of the action $A \equiv \int L d^3 x dt$ defined using equation (13) to zero for arbitrary variations leads to a set of four equations. One is the continuity equation (7) the three other are:

$$\frac{dv}{dt} = \frac{1}{2}\vec{v}^2 - w, \tag{17}$$

in which $w$ is the specific enthalpy.

$$\frac{d\alpha[\chi,\eta,v]}{dt} = \frac{\vec{\nabla}\eta \cdot \vec{J}}{\rho}, \tag{18}$$

$$\frac{d\beta[\chi,\eta,v]}{dt} = -\frac{\vec{\nabla}\chi \cdot \vec{J}}{\rho}. \tag{19}$$

In all the above equations and $\vec{v}$ is given by equation (11) and $\alpha,\beta$ are defined as:

$$\alpha = \frac{\vec{\nabla}\eta \cdot (\vec{B} \times (\vec{v} - \vec{\nabla}v))}{\vec{B}^2}$$

$$\beta = -\frac{\vec{\nabla}\chi \cdot (\vec{B} \times (\vec{v} - \vec{\nabla}v))}{\vec{B}^2}. \tag{20}$$

The above equations have been shown to be equivalent mathematically [2] to the standart formulation of magnetohydrodynamics.

**TOPOLOGICAL CONSTANTS OF MOTION**

Magnetohydrodynamics is known to have the following two topological constants of motion; one is the magnetic helicity:

$$H_M \equiv \int \vec{B} \cdot \vec{A} d^3 x, \tag{21}$$

which is known to measure the degree of knottiness of lines of the magnetic field $\vec{B}$ [20]. The domain of integration in equation (21) is the entire space, obviously regions containing a null magnetic field will have a null contribution to the integral. In the above equation $\vec{A}$ is the vector potential defined implicitly by the equation (1). The other topological constant is the magnetic cross helicity:

$$H_C \equiv \int \vec{B} \cdot \vec{v} d^3 x, \tag{22}$$

characterizing the degree of cross knottiness of the magnetic field and velocity lines. The domain of integration in equation (22) is the magnetohydrodynamic flow domain.

# REPRESENTATION IN TERMS OF THE MAGNETOHYDRODYNAMIC POTENTIALS

Let us write the topological constants given in equation (21) and equation (22) in terms of the magnetohydrodynamic potentials $\chi, \eta, \nu, \rho$ introduced in previous sections. First let us combine equation (1) with equation (10) to obtain the equation:

$$\vec{\nabla} \times (\vec{A} - \chi \vec{\nabla} \eta) = 0, \tag{23}$$

this leads immediately to the result:

$$\vec{A} = \chi \vec{\nabla} \eta + \vec{\nabla} \zeta, \tag{24}$$

in which $\zeta$ is some function. Let us now calculate the scalar product $\vec{B} \cdot \vec{A}$:

$$\vec{B} \cdot \vec{A} = (\vec{\nabla} \chi \times \vec{\nabla} \eta) \cdot \vec{\nabla} \zeta. \tag{25}$$

However, since we can define a local vector basis: $(\vec{\nabla} \chi, \vec{\nabla} \eta, \vec{\nabla} \mu)$ based on the magnetic field lines. In which in additon to $\chi, \eta$ we have added another coordinate the magnetic metage $\mu$ which paremtrize the distance along the magnetic field lines [1,13] we can write $\vec{\nabla} \zeta$ as:

$$\vec{\nabla} \zeta = \frac{\partial \zeta}{\partial \chi} \vec{\nabla} \chi + \frac{\partial \zeta}{\partial \mu} \vec{\nabla} \mu + \frac{\partial \zeta}{\partial \eta} \vec{\nabla} \eta. \tag{26}$$

Hence we can write:

$$\vec{B} \cdot \vec{A} = \frac{\partial \zeta}{\partial \mu} (\vec{\nabla} \chi \times \vec{\nabla} \eta) \cdot \vec{\nabla} \mu = \frac{\partial \zeta}{\partial \mu} \frac{\partial (\chi, \eta, \mu)}{\partial (x, y, z)}. \tag{27}$$

Let us think of the entire space outside the magnetohydrodynamic domain as containing low density matter in this case we can define the metage $\mu$ over the entire portion of space containing magnetic field lines and the integration domain of equation (21) and equation (22) coincide. Now we can insert equation (27) into equation (21) to obtain the expression:

$$H_M = \int \frac{\partial \zeta}{\partial \mu} d\mu d\chi d\eta. \tag{28}$$

The reader should notice that in some scenarios it may be that the flow domain should be divided into patches in which different definitions of $\mu, \chi, \eta$ apply to different domains, we do not see this as a limitation for our formalism since the topology of the flow is conserved by the equations of Magnetohydrodynamics. In those cases $H_M$ should be calculated as sum of the contributions from each patch. We can think about the magnetohydrodynamic domain as composed of thin closed tubes of magnetic lines each labelled by $(\chi, \eta)$. Performing the integration along such a thin tube in the metage direction results in:

$$\oint_{\chi, \eta} \frac{\partial \zeta}{\partial \mu} d\mu = [\zeta]_{\chi, \eta}, \tag{29}$$

in which $[\zeta]_{\chi, \eta}$ is the discontinuity of the function $\zeta$ along its cut. Thus a thin tube of magnetic lines in which $\zeta$ is single valued does not contribute to the magnetic helicity integral. Inserting equation (29) into equation (28) will result in:

$$H_M = \int [\zeta]_{\chi,\eta} d\chi d\eta = \int [\zeta] d\Phi, \qquad (30)$$

in which we have used $d\Phi = \vec{B} \cdot d\vec{S} = \vec{\nabla}\chi \times \vec{\nabla}\eta \cdot d\vec{S} = d\chi\, d\eta$. Hence:

$$[\zeta] = \frac{dH_M}{d\Phi}, \qquad (31)$$

the discontinuity of $\zeta$ is thus the density of magnetic helicity per unit of magnetic flux in a tube. We deduce that the Sakurai representation does not entail zero magnetic helicity, rather it is perfectly consistent with non zero magnetic helicity as was demonstrated above. Notice however, that the topological structure of the magnetohydrodynamic flow constrain the gauge freedom which is usually attributed to vector potential $\vec{A}$ and limits it to single valued functions. Moreover, while the choice of $\vec{A}$ is arbitrary since one can add to $\vec{A}$ an arbitrary gradient of a single valued function which may lead to different choices of $\zeta$ the discontinuity value $[\zeta]$ is not arbitrary and has a physical meaning given above. The main features of this novel "Magnetic Aharanov-Bohm effect" are simliar to the features of the standard Aharanov-Bohm effect.

1. A domain that is not simply connected, since the internal magnetic flux is knotted inside the external magnetic flux line (see figure 3).
2. The external magnetic field line does not touch the internal flux yet the $\zeta$ function is not single valued due to that line – non locality.
3. The potential vector field has a gradient of a non-single valued function part.
4. Gauge freedom is not gone but only limited to single-valued gauges.

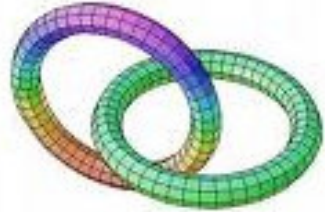

Figure **3**: Knotted magnetic field lines with none zero magnetic helicity and a non-single valued $\zeta$ .

Let us now introduce the velocity expression given in equation (11) and calculate the scalar product of $\vec{B}$ and $\vec{v}$, using the same arguments as in the previous paragraph will lead to the expression:

$$\vec{v} \cdot \vec{B} = \frac{\partial \nu}{\partial \mu}(\vec{\nabla}\chi \times \vec{\nabla}\eta) \cdot \vec{\nabla}\mu = \frac{\partial \nu}{\partial \mu} \frac{\partial(\chi,\eta,\mu)}{\partial(x,y,z)}. \qquad (32)$$

Inserting equation (32) into equation (22) will result in:

$$H_C = \int \frac{\partial \nu}{\partial \mu} d\mu d\chi d\eta. \qquad (33)$$

We can think about the magnetohydrodynamic domain as composed of thin closed tubes of magnetic lines each labelled by $(\chi,\eta)$. Performing the integration along such a thin tube in the metage direction results in:

$$\oint_{\chi,\eta} \frac{\partial \nu}{\partial \mu} d\mu = [\nu]_{\chi,\eta}, \qquad (34)$$

in which $[\nu]_{\chi,\eta}$ is the discontinuity of the function $\nu$ along its cut. Thus a thin tube of magnetic lines in which $\nu$ is single valued does not contribute to the cross helicity integral. Inserting equation (34) into equation (33) will result in:

$$H_C = \int [\nu]_{\chi,\eta} d\chi d\eta = \int [\nu] d\Phi. \qquad (35)$$

Hence:

$$[\nu] = \frac{dH_C}{d\Phi}, \qquad (36)$$

the discontinuity of $\nu$ is thus the density of cross helicity per unit of magnetic flux. We deduce that a flow with null cross helicity will have a single valued $\nu$ function alternatively, a non single valued $\nu$ will entail a non zero cross helicity. Furthermore, from equation (17) it is obvious that:

$$\frac{d[\nu]}{dt} = 0. \qquad (37)$$

We conclude that not only is the magnetic cross helicity conserved as an integral quantity of the entire magnetohydrodynamic domain but also the (local) density of cross helicity per unit of magnetic flux is a conserved quantity as well.

The main features of this novel "Cross Aharanov-Bohm effect" are simliar to the features of the standard Aharanov-Bohm effect:

1. A domain that is not simply connected, since the internal magnetic flux is knotted inside the external stream line.
2. The stream line does not touch the internal flux yet the $\nu$ function is not single valued due to that line – non locality.
3. The velocity field has a gradient of a non-single valued function part, this part is interpreted as a phase according to Bohm's causal interpretation correspondence see equation (4).

**CONCLUSION**

It is shown that there are two inherent Aharonov - Bohm effects in magnetohydrodymanics (MHD). In each case a magnetic flux induces a "phase" on quantities that do not come under the influence of the magnetic field directly. Those quantities include the velocity fields and "external" magnetic field. The phases $\varsigma, \nu$ quantify two well known Topological conservation laws of the magnetic and cross helicities. $\nu$ is useful for introducing a very efficient variational principle for MHD which is given in terms of only four independent functions for non-stationary flows. This is less than the seven variables which appear in the standard equations of MHD. More over the discontinuity $[\nu]$ is a conserved quantity along the MHD flow.